\documentclass[aps,prb,twocolumn,showpacs,superscriptaddress,floatfix]{revtex4-1}
\usepackage{amssymb,amsmath}
\usepackage{graphicx}
\usepackage{float}
\usepackage{bm}
\usepackage{multirow}
\usepackage{dcolumn}
\usepackage{tabularx}
\usepackage{longtable}
\usepackage[dvipsnames,usenames]{color}

\newlength{\dbarheight}

\begin{document}
%========================================================================= 
\title{Improper ferroelectricity and multiferroism in 2H-BaMnO$_3$}

\author{Julien Varignon}
\affiliation{Physique Th\'eorique des Mat\'eriaux, Universit\'e de Li\`ege
  (B5), B-4000 Li\`ege, Belgium}

\author{Philippe Ghosez}
\affiliation{Physique Th\'eorique des Mat\'eriaux, Universit\'e de Li\`ege
  (B5), B-4000 Li\`ege, Belgium}
\date{\today}
%========================================================================= 

\begin{abstract}  
  Using first-principles calculations, we study theoretically the stable 2H
  hexagonal structure of BaMnO$_3$. We show that from the stable high
  temperature $P6_3/mmc$ structure, the compound should exhibit an improper
  ferroelectric structural phase transition to a $P6_3cm$ ground
  state. Combined with its antiferromagnetic properties, 2H-BaMnO$_3$ is
  therefore expected to be multiferroic at low temperature. The phase
  transition mechanism in $\rm BaMnO_3$ appears similar to what was reported
  in YMnO$_3$ in spite of totally different atomic arrangement, cation sizes
  and Mn valence state.
\end{abstract}
\pacs{63.20.kk, 78.30.-j, 63.20.dk, 75.85.+t, 63.20.-e}
%========================================================================= 
\maketitle
%========================================================================= 

Multiferroic compounds combining ferroelectric and (anti-)ferromagnetic orders
present both fundamental and technological interests and have been the topic
of intensive researches during the last
decade~\cite{Kimura_Nature,Goto_PRL92,Hur_Nature429}. The identification of
new single-phase multiferroics has been particularly challenging.  In this
search, a special emphasis was placed on the family of multifunctional ABO$_3$
compounds within which belong many popular ferroelectric and magnetic
materials.  Unfortunately, few multiferroic ABO$_3$ oxides have been
identified to date. This scarcity was justified in terms of typical antagonist
metal $d$-state occupacy requirements~\cite{Hill00}: on the one hand,
ferroelectricity in typical compounds like BaTiO$_3$ arises from the
hybridization between occupied O $2p$ orbitals and empty Ti $3d$ states and
requires $d^0$-ness while, on the other hand, magnetism requires partial $d$
orbitals occupacy, so that ferroelectricity and magnetism often appear to be
mutually exclusive.

Exceptions to this rule exist however. In the antiferromagnetic CaMnO$_3$
perovskite, a ferroelelectric instability exists in spite of partial $d$-state
occupacy. In bulk samples, this instability is suppressed by
antiferrodistortive oxygen motions but it was predicted
theoretically~\cite{Bhattacharjee09}, and recently confirmed
experimentally~\cite{Gunter12}, that CaMnO$_3$ can be made ferroelectric, and
{\it de facto} multiferroic, under appropriate tensile epitaxial strain. In
the same spirit, it was proposed that, due to its larger volume, the
perovskite form of BaMnO$_3$ should develop a ferroelectric antiferromagnetic
ground state~\cite{Bhattacharjee09,Rondinelli09}.

\begin{figure}
\begin{center}
\begin{minipage}[l]{4cm}
 \resizebox{4cm}{!}{\includegraphics{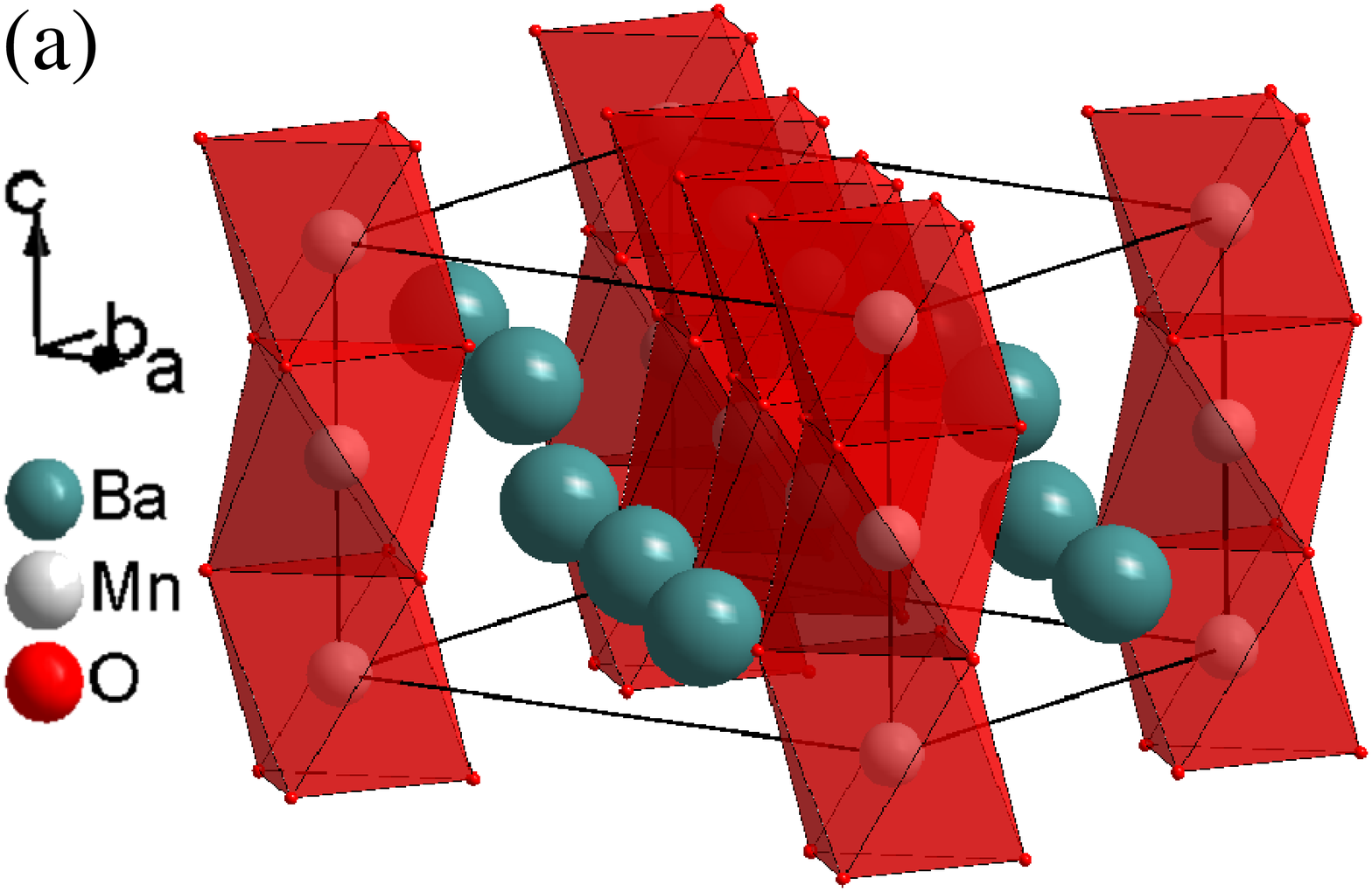}}
\end{minipage}
\hfill
\begin{minipage}[r]{4cm}
  \resizebox{4cm}{!}{\includegraphics{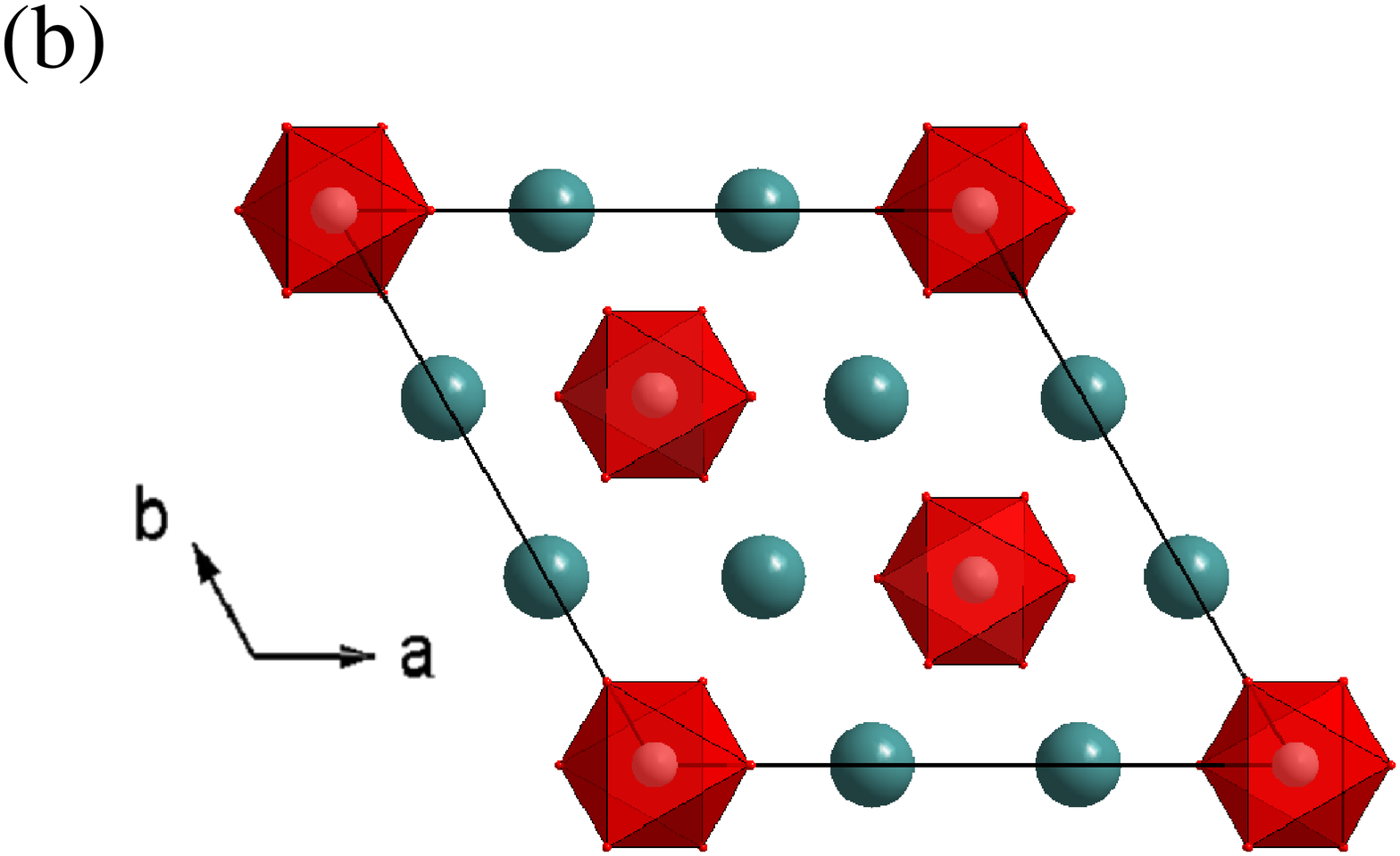}}
\end{minipage}
 \end{center}
 \caption{Crystallographic structure of $\rm 2H-BaMnO_3$ compound at low
   temperature ($P6_3cm$ space group). (a) Face-sharing $\rm MnO_6$ octahedra
   forming chains along the $\vec c$ axis.  (b) Triangular arrangement of the
   $\rm MnO_6$ octahedra in the $(\vec a, \vec b)$ plane.}
\label{f:BMO}
\end{figure}

BaMnO$_3$ does however not naturally crystallize in the perovskite structure
with corner-sharing oxygen octahedra. Due to its large Goldschmidt tolerance
factor, it stabilizes instead in a hexagonal 2H structure with face-sharing
oxygen octahedra (Figure \ref{f:BMO})~\cite{Hardy62,Chamberland70}. Increasing
pressure decreases the hexagonal character, yielding at high pressure various
possible forms including a 4H hexagonal structure combining corner- and
face-sharing octahedra, which is intermediate between 2H and cubic perovskite
structures~\cite{Chamberland70}. The stability of the 2H structure was
confirmed from first-principles calculations~\cite{Sondena07} but its exact
symmetry remains under debate.

%\begin{figure}[h!]

From the similarity of spectra with BaNiO$_3$~\cite{Lander51}, Hardy
originally assigned the 2H phase of BaMnO$_3$ at room temperature to the polar
$P6_3mc$ space group~\cite{Hardy62}. Combined Raman and infra-red phonon
measurements performed by Roy and Budhani~\cite{Roy98} are compatible with
such a polar character, that is also in line with a recent report of room
temperature ferroelectricity in $\rm 2H-BaMnO_3$ by Satapathy et
al~\cite{Satapahthy12}.
  
Nevertheless, Christensen and Ollivier reported that 2H-BaMnO$_3$ should
better belong to the $P6_3/mmc$ space group~\cite{Christensen72}. We notice
that a similar re-assignment to $P6_3/mmc$ was also proposed for
BaNiO$_3$~\cite{Takeda76} that had served as reference material to Hardy.
This was also the choice of Cussen and Battle~\cite{Cussen00} who found no
clear evidence to assign the room temperature structure of 2H-BaMnO$_3$ to the
polar $P6_3mc$ space group but instead chose the non-polar $P6_3/mmc$ space
group. Going further, they highlighted a structural phase transition between
room temperature and 80 K to a polar $P6_3cm$ space group, yielding unit-cell
tripling. Moreover, while Christensen and Ollivier~\cite{Christensen72} were
not observing any clear magnetic order above 2.4 K, Cussen and
Battle~\cite{Cussen00} reported an antiferromagnetic ordering below $T_N=59K$
in the $P6_3cm$ phase, with neighboring spins antiferromagnetically coupled
along the polar axis and forming a triangular arrangement perpendicular to it.
Although it was not explicitly emphasized, according to this, BaMnO$_3$ should
be multiferroic at low temperature in its stable 2H polymorph. Nevertheless
this has to be taken with caution since
%, although they reported a $P6_3/mmc$ space group at room temperature, 
Cussen and Battle cannot rule out the possibility of a centrosymmetric
$P\bar{3}c1$ space group (subgroup of $P6_3/mcm$) at low
temperature~\cite{Cussen00}. Also, no consensus has been achieved yet
regarding the stable structure at room temperature~\cite{Satapahthy12}.

In order to get additional insight into the possible ferroelectric nature of
$\rm 2H-BaMnO_3$, we performed first-principles calculations. We show that the
high-temperature phase should belong to the $P6_3/mmc$ space group and that
the compound should undergo an improper ferroelectric structural phase
transition to a $P6_3cm$ phase, through a mechanism similar to what happens in
hexagonal YMnO$_3$. Combined with its anti-ferromagnetic properties, this
study confirms that $\rm 2H-BaMnO_3$ should be multiferroic at low
temperature.

Our calculations have been performed within density functional theory (DFT)
using the CRYSTAL09 package~\cite{CRYSTAL09}. We used the B1-WC hybrid
functional~\cite{B1WC} that was previously found appropriate for the study of
multiferroic oxides~\cite{Alina}. Relativistic pseudopotentials and a $3\zeta$
valence basis sets were used for the barium atoms~\cite{baseBa}.  Small core
Hay-Wadt pseudopotentials, associated with $2\zeta$ valence basis
sets~\cite{baseMn}, were used for the manganese atoms. All-electron $2\zeta$
basis sets, specifically optimized for $O^{2-}$, were used for the oxygen
atoms~\cite{baseO}. The calculations were carried out on a $4\times 4 \times
8$ $k$-point grid for the tripled unit cell (30 atoms). The polarization was
computed from the displacement of Wannier function
centers~\cite{Vanderbilt93,Zicovich2001}. We worked within a collinear-spin framework
assuming an A-type antiferromagnetic order with spins antiparallel along the
$\vec c$ direction and aligned in the $(\vec a, \vec b)$ plane. Calculations
with a ferromagnetic ordering provided similar results. We performed geometry
optimization of atomic positions until the root mean square values of atomic
forces were lower than $5\times 10^{-5}\,\rm Ha.Bohr^{-1}$ and until the root
mean square of atomic displacements were lower than $5\times 10^{-5}\,\rm
Bohr$.  For better comparison with the experiment, results reported below have
been obtained at fixed experimental lattice parameters as reported by Cussen
and Battle at room temperature~\cite{Cussen00}.  Similar results have been
obtained adopting the lattice parameters reported at 1.7 K and 80 K. Full
structural optimization including lattice parameters were also performed
providing qualitatively similar results but yielding a smaller ferroelectric
distortion, due to small theoretical errors on the lattice parameters.
  
First, we performed a structural optimization of $\rm 2H-BaMnO_3$ starting
from a polar $P6_3mc$ configuration as initially proposed by
Hardy~\cite{Hardy62}.  Surprisingly, the system went back to a $P6_3/mmc$
configuration, with atomic positions (one degree of freedom, $x_O = 0.1473$)
in good agreement with what was reported by Cussen and Battle ($x_O =
0.1495$)~\cite{Cussen00}. This demonstrates theoretically that the system does
not want a priori to be ferroelectrically distorted but prefers to adopt a
non-polar $P6_3/mmc$ symmetry as first proposed by Christensen and Ollivier.
This is in conflict with the recent claim of a room temperature $P6_3mc$
ferroelectric phase by Satapathy {\it et al}~\cite{Satapahthy12}. However the
crystal structure they report corresponds in fact to a $P6_3/mmc$
phase~\cite{footnote1} and is therefore inconsistent with the polarization
they measure.  In order to further confirm our result, we computed the
zone-center phonon modes of the relaxed $P6_3/mmc$ phase without identifying
any zone-center unstable mode, the lowest polar $\Gamma_2^-$ mode being at a
frequency around $56\,cm^{-1}$.
  
In order to test the possibility of a phase transition to a $P6_3cm$ or a
$P\bar{3}c1$ phase, we computed the phonons at the zone-boundary point q=(1/3,
1/3, 0). In this case, we identified an unstable mode at 25$i$ cm$^{-1}$ of
$K_3$ symmetry. This mode is related to oxygen and manganese motions along the
polar axis. Its condensation produces a unit-cell tripling and brings the
system into the $P6_3cm$ symmetry.  We notice that no unstable mode of $K_1$
symmetry is identified, rulling out the possibility of a transition to a phase
of $P6_3/mcm$ or $P\bar{3}c1$ symmetry.

%\begin{table}[h!]
\begin{table}
\begin{center}
\begin{tabular}{lccc}
\hline
\hline
         &Theory &Exp.~\cite{Cussen00} &Exp.~\cite{Cussen00} \\
         &(present) &(1.7K) &(80K) \\
\hline
$x_{Ba}$      & 0.334    & 0.339 & 0.332 \\
$z_{Ba}$      & 0.208   & 0.230 & 0.238 \\
$z_{Mn_1}$ &0.000     & 0.000 & 0.000 \\
$z_{Mn_2}$ &-0.067   & -0.048 & -0.037 \\
$x_{O_1}$   &0.147    & 0.149 & 0.150 \\
$z_{O_1}$   &0.251    & 0.248 & 0.250  \\
$x_{O_2}$   &0.667    & 0.664 & 0.667 \\
$y_{O_2}$   &0.481    &0.482 & 0.483 \\
$z_{O_2}$  &0.182     & 0.200 & 0.212 \\
\hline
\hline
\end{tabular}
\end{center}
\caption{Atomic positions (in fractional coordinates)  in the $P6_3cm$ structure of 2H-BaMnO$_3$. 
  Experimental values are those of Cussen and Battle at 1.7K and 80K~\cite{Cussen00}.}
\label{t:opt}
\end{table}

\begin{table}
\begin{center}
\begin{tabular}{lccccc}
\hline
\hline
                &  $\Gamma_1^+$ & $\Gamma_2^-$ & $K_1$ & $K_3$ & total \\
\hline
Exp. (1.7K)       &0.015 &0.144 &0.154 &0.534 &0.574 \\
Exp. (80K)        &0.035 &0.139 &0.035 &0.420 &0.445 \\ 
Theory             &0.001 &0.034 &0.005 &0.766 &0.767  \\
%Theory (RC, AFM)  &0.004 &0.011 &0.000 &0.160 &0.161 \\  
%Theory (FC, FM) &0.004 &0.007 &0.000 &0.611 &0.611 \\
\hline
\hline
\end{tabular}
\end{center}
\caption{Amplitude (in $\rm \AA$) of symmetry allowed displacements for
  experimental and theoretical structures (at fixed lattice parameters).}
\label{t:irreps}
\end{table}

Atomic relaxation of the $P6_3cm$ phase has been performed, yielding a gain of
energy of 21 meV/f.u. respect to the $P6_3/mmc$ structure. The relaxed atomic
positions are reported in Table \ref{t:opt} and show qualitative agreement
with experimental data.  BaMnO$_3$ appears in our hybrid functional
calculation as an insulator with an electronic bandgap of 4.15 eV. As allowed
by the polar nature of the $P6_3cm$ space group, this phase also develops a
sizable spontaneous polarization $P_s=0.92\,\mu C.cm^{-2}$ along the polar
axis. Since this polarization can be either up or down (and is so {\it a
  priori} switchable), this phase can be labelled as ferroelectric.

Interestingly, although the condensation of the $K_3$ mode breaks inversion
symmetry, this mode is non-polar (i.e. its mode effective charge is zero) and
its condensation alone does not induce any significant
polarization~\cite{footnote2}. Further insight into the mechanism yielding a
significant polarization at the phase transition from $P6_3/mmc$ to $P6_3cm$
is therefore needed and can be obtained from a decomposition of the atomic
distortion condensed at the transition into symmetry adapted modes, as
achievable with the {\it Amplimodes} software of the {\it Bilbao
  Crystallographic server} ~\cite{Bilbao1,Bilbao2}.

The results of this decomposition are reported in Table \ref{t:irreps}. Both
theory and experiment find the structural distortion to be principally
dominated by a $K_3$ character. Then, and although underestimating the
experiment, another significant atomic motion observed theoretically has a
$\Gamma_2^-$ character. As further discussed below, it is in fact this
additional polar distortion that will produce a sizable polarization. We
notice that the appearance of additional $\Gamma_1^+$ and $K_1$ atomic
distortions are also possible by symmetry but are not observed in our
calculations.  This last result contrasts with experimental data that suggest
the presence of a sizable $K_1$ contribution. We notice however that
experimental data at 1.7K and 80K are not fully compatible in that respect, so
that further refined experimental characterization of this phase would likely
be very useful to clarify that issue.

The interplay between dominant $K_3$ and $\Gamma_2^-$ atomic distortions can
be better understood from an expansion of the energy around the $P6_3/mmc$
phase in terms of their amplitude $\mathcal{Q}_{K_3}$ and
$\mathcal{Q}_{\Gamma_2^-}$ :

\begin{eqnarray}
  \mathcal{F}(\mathcal{Q}_{K_3},\mathcal{Q}_{\Gamma_2^-}) & = & \alpha_{20}
  \mathcal{Q}_{K_3}^2 + \alpha_{02}\mathcal{Q}_{\Gamma_2^-}^2
  +\beta_{40}\mathcal{Q}_{K_3}^4 + \beta_{04}\mathcal{Q}_{\Gamma_2^-}^4
  \nonumber \\
  & & +\beta_{31} \mathcal{Q}_{K_3}^3\mathcal{Q}_{\Gamma_2^-} + \beta_{22}
  \mathcal{Q}_{K_3}^2\mathcal{Q}_{\Gamma_2^-}^2    \label{e:expansion}
\end{eqnarray}

\begin{figure}
\centering
\resizebox{8cm}{!}{\includegraphics{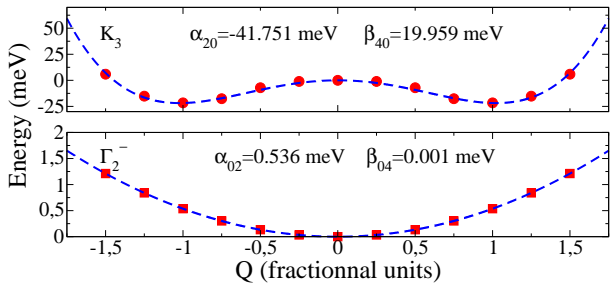}}
\caption{Evolution of the energy (in meV) of 2H-BaMnO$_3$ in terms of
  the amplitude $\mathcal{Q}_{K_3}$ and $\mathcal{Q}_{\Gamma_2^-}$ (in
  fractional units) of the $K_3$ and $\Gamma_2^-$ modes. The
  $P6_3/mmc$ structure was taken as reference.}
\label{f:K3}
\end{figure}

The coefficients of this energy expansion were determined from DFT
calculations on a $\sqrt{3} \vec a_{PE} \times \sqrt{3} \vec b_{PE}
\times \vec c_{PE}$ optimized $P6_3/mmc$ supercell with various
amplitudes of $\mathcal{Q}_{K_3}$ and $\mathcal{Q}_{\Gamma_2^-}$
condensed. The results are summarized in Fig.~\ref{f:K3} and are
consistent with what was previously discussed. The $\Gamma_2^-$ polar
mode is stable and associated to a single well while the $K_3$ mode is
unstable and has a typical double-well shape.

%\begin{figure}[h!]

\begin{figure}[h!]
\centering
\resizebox{8cm}{4cm}{\includegraphics{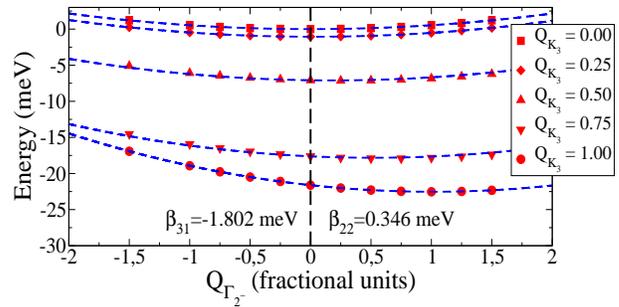}}
\caption{Energy (in meV) as a function of $\Gamma_2^-$ allowed
  displacements  at fixed $\mathcal{Q}_{K_3}$ (in fractional units).}
\label{f:Gm2}
\end{figure}

The phase transition of 2H-BaMnO$_3$ from $P6_3/mmc$ to $P6_3cm$ is therefore
clearly driven by the unstable $K_3$ primary mode. Then the sizable
polarization arises from the additional condensation of the stable polar
$\Gamma_2^-$ secondary mode which, as further illustrated in Fig. \ref{f:Gm2},
results from a progressive shift of the minimum of the single well, produced
by $\mathcal{Q}_{K_3}$ through the $\beta_{31} \mathcal{Q}_{K_3}^3
\mathcal{Q}_{\Gamma_2^-}$ coupling term.

\begin{figure}
\begin{center}
\resizebox{8cm}{!}{\includegraphics{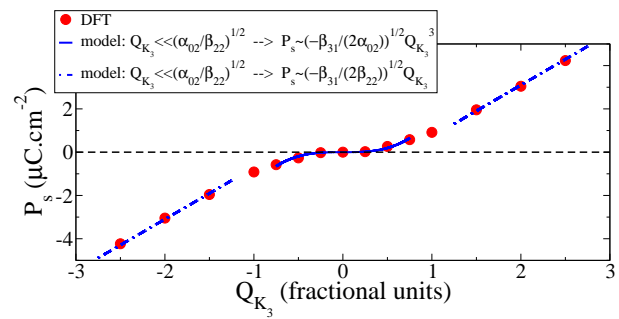}}
\end{center}
\caption{Spontaneous polarization (in $\mu C.cm^{-2}$) as a function of the
  primary order parameter $\mathcal{Q}_{K_3}$ (in fractional units).}
\label{f:PK3}
\end{figure}

This mechanism is similar to what was reported by Fennie and Rabe
~\cite{Fennie_RapidComm} in YMnO$_3$ and allows us to identify 2H-BaMnO$_3$ as
an {\it improper} ferroelectric. The evolution of the polarization with
$\mathcal{Q}_{K_3}$ is represented in Fig. \ref{f:PK3}. We first notice that
the polarization estimated within the model from the amplitude of
$\mathcal{Q}_{\Gamma_2^-}$ assuming constant Born effective charges
($Z^*_{Ba}= +2.10 e$, $Z^*_{Mn}= +6.45 e$ and $Z^*_{O}= -2.85 e$ along the
$\vec c$ axis in the $P6_3/mmc$ phase \cite{Note-Z}) properly reproduces the
DFT calculation. As discussed by Fennie and Rabe ~\cite{Fennie_RapidComm}, two
independent regimes can be identified: at small $\mathcal{Q}_{K_3}$ amplitude,
$\mathcal{Q}_{\Gamma_2^-}$ (and therefore $P_s$) evolves like
$\mathcal{Q}_{K_3}^3$ while at larger amplitude it evolves linearly.  Contrary
to YMnO$_3$ that has its ground state in the linear regime, BaMnO$_3$ is still
in the small amplitude regime in our calculation.

Although they both crystallize in the same hexagonal $P6_3/mmc$ space group at
high-temperature, the similarity between BaMnO$_3$ and YMnO$_3$ is astonishing.
First, the valence state of the cations are different in both compounds
($+2/+4$ in BaMnO$_3$ versus $+3/+3$ in YMnO$_3$). The cation sizes are also
distinct in both compounds yielding a Goldschmidt tolerance factor $t>1$ in
BaMnO$_3$ and $t<1$ in YMnO$_3$. Consequently, the atomic structure is totally
different in both compounds: the Mn atom is surrounded by face-sharing oxygen
octahedra in BaMnO$_3$ and corner-sharing oxygen trigonal bipyramids in
YMnO$_3$.  The atomic motion associated to the unstable $K_3$ mode is also
rather different. In $\rm YMnO_3$, $K_3$ motions correspond to a tilt of $\rm
MnO_5$ bipyramids accompanied by a shift of Y atoms and O atoms forming the
basis of $\rm MnO_5$ bipyramids along the polar axis. This leads to a more
symmetric environment for the Mn atom, reducing its oxidation degree and
making the charge of all O atoms similar. In BaMnO$_3$, the $K_3$ mode is
associated only to a shift of Mn and O atoms along the polar axis, increasing
the absolute value of the oxidation degree of all atoms. The fact that both
compounds exhibit an unstable $K_3$ mode, that can drive improper
ferroelectricity appears therefore rather fortuitous.

Finally, we studied the magnetic properties of ground state $\rm 2H-BaMnO_3$.
In their seminal work, Christensen and Ollivier~\cite{Christensen72} did not
identify any clear magnetic order above 2.4 K but more recently, Cussen and
Battle~\cite{Cussen00} reported an antiferromagnetic ordering below $T_N=59K$
in the $P6_3cm$ phase: they proposed a configuration comparable to YMnO$_3$ in
which neighboring spins are antiferromagnetically aligned along the polar axis
and forming a triangular arrangement perpendicular to it.

Although working within a collinear-spin approximation, we estimated the
coupling constants $J_{ij}$ between spins at site $i$ and $j$ in the $P6_3cm$
phase through the fit of the Heisenberg effective Hamiltonian limited to
nearest-neighbor interactions: $H_{Heisenberg}= - \sum_{i<j}J_{ij}\vec
S_i\cdot \vec S_j$.  2H-BaMnO$_3$ presents two distinct magnetic interactions:
one along the polar axis and one perpendicular to it. We found a strong
antiferromagnetic coupling $J_c=-32.77$~meV along the polar axis and a much
weaker interchain antiferromagnetic coupling $J_{ab}=-0.14$~meV in plane,
comparable to that computed in YMnO$_3$~\cite{JYMO_hybride}. The estimate of
$J_{ab}$ has certainly to be taken with more caution than $J_c$: in YMnO$_3$,
collinear-spin calculations significantly underestimated the magnitude of
$J_{ab}$ on a similar triangular arrangement of spins~\cite{JYMO_hybride}
(0.59~meV instead of 2.3-3~meV experimentally). Nevertheless, a strong
anisotropy of the magnetic interactions seems to be likely and can be
explained from simple geometry arguments. Along the polar axis, there are
face-sharing octahedra and the super-exchange term between two neighboring Mn
atoms goes through three different Mn-O-Mn exchange paths, increasing the
overlap between Mn $3d$ and O $2p$ orbitals and the magnetic coupling. On the
other hand, Mn-Mn interchain distances are longer and the super-exchange term
in-plane has to go consecutively through O and Ba atoms (see
Fig.~\ref{f:BMO}); the overlap between orbitals will be smaller and the
magnetic exchange interactions weaker. So, although it is proposed that both
share the same magnetic arrangement, the antiferromagnetic coupling in
2H-BaMnO$_3$ appears substantially different than in YMnO$_3$ for which the
strongest magnetic interactions are in-plane.

%\section{conclusion}

In summary, we propose in this Letter that 2H-BaMnO$_3$ is an {\it improper} 
ferroelectric. Our calculations moreover confirms its antiferromagnetic character, 
making it a multiferroic compound amazingly similar to YMnO$_3$, in spite of a radically 
different atomic arrangement.  Improper ferroelectrics have recently generated 
increasing interest in view of their unusual electrical properties \cite{Stengel12}. 
Not so many ABO$_3$ improper ferroelectrics have been reported yet and our study 
illustrates that such behavior can also happen in hexagonal 2H compounds. We 
hope that our theoretical work will motivate the search of new improper ferroelectrics 
and multiferroics in this class of compounds.

\acknowledgments
This work was supported by  the European project OxIDes (FP7) and the ARC 
project TheMoTher. PhG acknowledges Research Professorship from the Francqui 
foundation. 

%%%%%%%%%%%%%%%%%%%%%%%%%%%%%%%%%%%%%%%%%%%%%%%%%%%%%%%%%%%%%%%%%%%%%%%%%


\begin{thebibliography}{}
\bibitem{footnote1} In Table 1 of Ref.~\onlinecite{Satapahthy12}, Ba
  and Mn positions were inverted. Then shifting back the z/c coordinates
   by $z_{Mn}$, we recover atomic position consistent with a non-polar 
   $P6_3/mmc$ phase within the experimental accuracy (i.e. the deviation 
   is extremely small and totally incompatible with the large polarization they report).
\bibitem{footnote2}
Condensation of the non-polar $K_3$ mode alone does not give rise to any 
sizable polarization : we get $P_s=0.01$ $\mu$C.cm$^{-2}$ as a result of the 
purely electronic response.
\bibitem{Note-Z}
Although still anomalous (and larger than in YMnO$3$), the $Z^*_{Mn}$ and
$Z^*_{O}$ are significantly smaller in the $P6_3/mmc$ phase than in the cubic 
perovskite structure ($Z^*_{Ba}= +1.89 e$, $Z^*_{Mn}= +7.97 e$ and 
$Z^*_{O}= -1.55 e/-6.72 e$ at the relaxed $a_0=3.88$ \AA)).
\end{thebibliography}
\end{document}